\documentclass[11pt]{article}
\usepackage{amssymb}
\usepackage{makeidx}
\usepackage{graphicx}
\usepackage{array}
\def\be{\begin{equation}}
\def\ee{\end{equation}}
\def\bea{\begin{eqnarray}}
\def\eea{\end{eqnarray}}
\def\ci{\cite}

\def\p{\phi}
\def\rp{\rho_\phi}

\begin{document}
\title{Cosmology and Quantum Field Theory: A study of the Nambu-Jona-Lasinio Model}\medskip
\author{Leonardo Quintanar G., Axel de la Macorra \\}
\date{}
\maketitle
\begin{center}Instituto de F\'{\i}sica,\\ Universidad Nacional Aut\'onoma de M\'exico\end{center}
\vspace*{7mm}
\begin{abstract}
\noindent
We review the Nambu-Jona-Lasinio model (NJL), proposed long time ago, as a four-fermion interaction theory with chiral symmetry. The theory is not renormalizable and presents a symmetry breaking due to quantum corrections which depends on the strength of the coupling constant. We may associate a phase transition with this symmetry breaking, leading from a fermion states to a fermion condensate, which can be described effectively by a scalar field. We are interested in this paper in the cosmological dynamics of the NJL model, and in the possibility that it can be related to  dark energy and/or dark matter, which form up to $95\% $ of the energy content of the universe at present time. We consider exclusively gravitational interaction between the  NJL and the SM particles.

\end{abstract}

\section{Introduction}
In the last years the study of our universe has received a great deal of attention since on the one hand fundamental theoretical cosmological questions remain unanswered and on the other hand we have now the opportunity to measure the cosmological parameters with an extraordinary precision. In the last decades, research in cosmology has revealed the presence of unexplained forms of matter and energy called Dark Energy "DE" and Dark matter "DM" making up to $95\% $ of the energy content of the universe at present time. The existence of DE was established with the study of supernovas SNIa, showing that the universe in not only expanding, but besides it is accelerating \cite{SN.1,SN.2}. Such behaviour can be explained by the existence of a new form of Dark Energy with a "strange" anti-gravitational property, which can be explain by a fluid with negative pressure called Dark Energy. Besides the SNIa
Dark Matter (DM) and Dark Energy (DE), is provided through the analysis of the Cosmic Microwave Background Radiation (CMB) \cite{CMB}, which has been measured by satellite WMAP \cite{WMAP}, and more recently by Planck mission \cite{PLANCK}, and  the dynamics of galaxies, clusters and super clusters, and  the study of the formation of Large Scale Structure \cite{LSS.1} in the universe and weak lensing (the gravitational deviation of light), and point out the existence of matter that do interacts  with ordinary standard model matter only weakly, as due to  gravity. Other measurement such as Baryon Accoustic Oscilalations "BAO" \ci{BAO.1,BAO.2}.\\
It has been established that our universe is flat and dominated at present time by Dark Energy "DE" and Dark Matter "DM" with
$\Omega_{DE}\simeq 0.692\pm 0.02$,  $\Omega_{m}= 0.308 \pm 0.009 $ and Hubble constant $H_o=(67.27\pm 0.66) km s^-1 Mpc^-1$ \ci{PLANCK}. However, the nature and dynamics of Dark Energy and Dark Matter is a topic of major interest
in the field \ci{DE.rev}. The equation of state "EOS"  of DE is at present
time $w_o \simeq -0.93\pm 0.13$ but we do not have a precise measurements of $w(z)$ as a function
of redshift  $z$ \ci{LSS.2,PLANCK}. Since the properties of Dark Energy are still under investigation, different DE parametrization have been proposed to help discern on the dynamics of DE \ci{DE.rev}-\ci{quint.ax}. Some of these  DE parametrization have the advantage of having a reduced number of parameters, but they may lack of a physical motivation
and may also be too restrictive.  Perhaps the best physically motivated candidates for Dark Energy are scalar fields which can be minimally coupled, only via gravity,  to other fluids \ci{SF,tracker, quint.ax} or can interact weakly in interacting Dark Energy "IDE"\ci{IDE,IDE.ax}. Scalar fields have been widely studied in the literature \ci{SF,tracker, quint.ax} and special interest was devoted to tracker fields \ci{tracker} since in this case the behavior of the scalar field $\phi$ is very weakly dependent on the initial conditions at a very early epoch and well before matter-radiation equality. In this class of models the fundamental question of why DE is relevant now, also called the coincidence problem, can be ameliorated by the insensitivity of the late time dynamics on the initial conditions of $\p$.\medskip\\
Nowadays there is a huge amount of ideas aimed to explain these unknown cosmological fluids from the theoretical point of view, none of them being still conclusive. This situation support and motivate our research. Given that our most successful theory of matter, the Standard Model of particle Physics (SM), is settled within the theoretical frame of Quantum Field Theory (QFT), it would be reasonable to ask a theory attempting to describe Dark fluids to be based on QFT as well. In this paper we study a fermion interaction theory with a chiral symmetry, the Nambu-Jona-Lasinio (NJL) model \ci{NJL}. Though this is an old and well known model in the context of hadron physics, it has interesting properties and it is worth to consider it with a different perspective, by studying its possible relevance for Cosmological Physics.  Other examples of QFT models of DE and DM have been proposed using gauge groups, similar to QCD in particle physics, and have been studied to understand the nature of Dark Energy \ci{GDE.ax} and also Dark Matter \ci{GDM.ax}.\medskip\\
We organized the present work as follows: In section (\ref{secnjl}) we present the NJL model. In section (\ref{cosmo}) we review the pertinent cosmological theory. Sections (\ref{dyn.weak}) and (\ref{dyn.strong}) presents a study of the cosmological dynamics of a NJL fluid with a weak and strong coupling, respectively. In section (\ref{coscte}) we consider the addition of a cosmological constant to our NJL fluid, and analyze the different possible behaviours. Finally, in section (\ref{conclu}) we summarize our results and present  conclusions.

\section{The Nambu-Jona-Lasinio model.}\label{secnjl}
Inspired by a, by then recently explained phenomenon in Superconductivity research, professors Y. Nambu and Jona-Lasinio, suggested that the mass of fermion particles (described by a Dirac equation) could be generated from a primary self four  fermions interaction, leading to a chiral symmetry  breaking \ci{NJL}. The proposed Lagrangian, invariant under chiral transformations, has the form \ci{NJL}
\begin{equation}
{\cal L}=i\bar\psi \gamma^{\mu}\partial_{\mu}\psi +\frac{g^2}{2}\left[(\bar\psi\psi)^2-(\bar\psi\gamma_5\psi)^2\right] ,
\label{Lnjl}
\end{equation}
where $\psi$ is a four-component spinor, and $g$ is a coupling constant. The interaction term
\begin{equation}
{\cal L}_{int}=\frac{g^2}{2} \left[(\bar\psi\psi)^2-(\bar\psi\gamma_5\psi)^2\right]
\label{Lnjleq1}\end{equation}
with no original mass term for the fermions.
As this coupling has dimension $-2$ in mass units, the theory is non-renormalizable. However, we are interested in considering the NJL model as an \textit{effective theory}, useful below certain energy scale. The theory (\ref{Lnjl}) describes a four-fermion interaction which can be expanded following conventional perturbation theory, and represented by Feynman diagrams:
\begin{figure}[h]\centering \includegraphics[scale=.5]{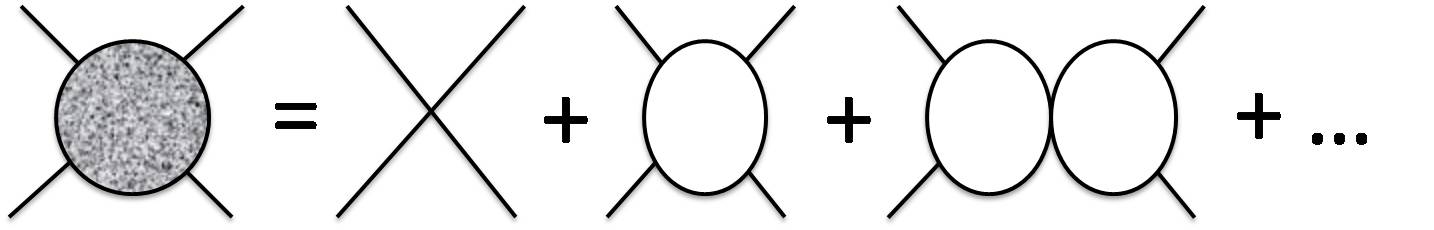}
\caption{Feynman diagram for a four-fermion interaction.} \label{fig00}
\end{figure}\\
The infinite number of fermion loops can be resumed giving a non-perturbative potential. This can be easily done by introducing an auxiliary scalar field $\phi$, and an equivalent lagrangian for eq.(\ref{Lnjleq1}) may be written in the form
\begin{equation}
{\cal L}_{int}=\lambda\phi\bar\psi\psi-\frac{1}{2}m^2\phi^2 .
\label{Lnjleq}
\end{equation}
The coefficient $\lambda$ is a Lagrange multiplier to be determined below, whereas $m$ is a mass dimension parameter introduced to make physical units consistent.
The term $\bar{\psi}\gamma_5\psi$ in eq.(\ref{Lnjl}) represents a pseudo-scalar quantity, and we have allowed ourselves to ignore the field contribution associated with it in the new lagrangian in eq.(\ref{Lnjleq}), as we would like to start to study the simplest possible model.
Using the Euler-Lagrange equations, $\partial_{\mu}\frac{\delta{\cal L}}{\delta(\partial_{\mu}\phi)}-\frac{\delta{\cal L}}{\delta\phi}=0$ for the field $\phi$, we find
\begin{equation}
\phi=\frac{g}{m}\bar\psi\psi , \label{phi}
\end{equation}
and clearly then eq.(\ref{Lnjleq1}) is equivalent to  eq.(\ref{Lnjleq}), neglecting the term $\bar{\psi}\gamma_5\psi$, if we identify  the lagrange multiplier as $\lambda=mg$.  Thus, we may read the fermion mass and the tree level scalar potential  $V_0$,  and we have respectively:
\begin{equation}
m_\psi^2=(mg\phi)^2,\quad V_0=\frac{1}{2}m^2\phi^2 . \label{pot0}
\end{equation}
The effect of quantum processes (represented by loop diagrams) may be taken into account through the well known Coleman-Weinberg potential $V_1=-\frac{1}{8\pi}\int p^2\log(p^2+m_{\psi}^2)d^2p$ (the minus sign is due to the fact that we are dealing with a fermionic field). This integral grows up indefinitely as the upper limit goes to infinity, i.e. it has an ultraviolet divergence. Because of the non-renormalizability of the theory, we can not avoid this divergence, so we regularize by introducing a cut-off $\Lambda$. This parameter define the energy scale below of which the theory is valid. We  define the $x$ variable as
\be
x\equiv \frac{m_\psi^2}{\Lambda^2} = {m^2g^2\phi^2\over \Lambda^2},\label{xphi}
\ee
and the potential becomes
\begin{eqnarray}
V_0&&= \frac{ \Lambda^2 x}{2g^2} ,\\
V_1&&=
-\frac{\Lambda^4}{16\pi^2}\left[x+x^2\log\left(\frac{x}{1+x}\right)+\log(1+x)\right].
\end{eqnarray}
For the seek of concision we also define
\begin{equation}
A\equiv \frac{\Lambda^4}{16\pi^2},\quad
 f(x)=x+x^2\log\left(\frac{x}{1+x}\right)+\log(1+x).\label{Af}
\end{equation}
In this way, taking quantum corrections into account we obtain an effective potential given by
\begin{equation}
V=V_0 +V_1 =\frac{\Lambda^2 x}{2g^2}-Af(x),
\end{equation}
with the complete potential
\begin{equation}
V=V_0+V_1=\frac{\Lambda^2 x}{2g^2}\left(1-\frac{g^2\Lambda^2}{8\pi^2}\right)-\frac{\Lambda^4}{16\pi^2} \left[x^2\log\left(\frac{x}{1+x}\right)+\log(1+x)\right].\\
\end{equation}
As a function of $\phi$ it can be written explicitly as
\begin{equation}
V(\phi)=\frac{1}{2}m^2\phi^2 -\frac{\Lambda^4}{16\pi^2}
\left[
\left({mg\phi\over \Lambda}\right)^2+
\left({mg\phi\over \Lambda}\right)^4\log
\left(\frac{\left({mg\phi\over \Lambda}\right)^2}{1+\left({mg\phi\over \Lambda}\right)^2}\right)
+\log\left(1+\left({mg\phi\over \Lambda}\right)^2\right)
\right] .\label{vefphi}
\end{equation}
Eq. (\ref{vefphi}) gives the complete NJL scalar potential, and we are interested in studying its cosmological implications.
Let us determine the asymptotic behaviour of the scalar potential $V$ in eq.(\ref{vefphi}).
To analyze the potential we seek for extremum points. For the funcion $f(x)$ in eq. (\ref{Af}) we have the derivative
\begin{equation}
\frac{df(x)}{dx}=2\left[1+x\log\left(\frac{x}{1+x}\right)\right],\label{df}
\end{equation}
and for the derivative of $V$ we have
\begin{eqnarray}
\frac{\partial V}{\partial \phi}&&=
\frac{m^2\Lambda^2\phi}{4\pi^2}\left\lbrace \frac{4\pi^2}{g^2\Lambda^2}-1 - (\frac{mg\phi}{\Lambda})^2\log\left( \frac{(\frac{mg\phi}{\Lambda})^2}{1+(\frac{mg\phi}{\Lambda})^2} \right)    \right\rbrace , \\
\frac{\partial V}{\partial \phi}&&=\frac{m^2\Lambda^2\phi}{4\pi^2}\left\lbrace \frac{4\pi^2}{g^2\Lambda^2}-1 -x\log\left( \frac{x}{1+x} \right)  \right\rbrace .
\end{eqnarray}
The condition $\frac{\partial V}{\partial \phi}=0$ implies the following equations:
\begin{equation}
i)\;\;\phi=0,\quad \mbox{ or }\quad ii)\;\; \frac{4\pi^2}{g^2\Lambda^2}-1= x\log\left(\frac{x}{1+x}\right). \label{dV1}
\label{dv1}\end{equation}
The first one says that the origin $\phi=0$ is an extremum, and if we take the second derivative $\frac{\partial^2 V}{\partial \phi^2}$
\begin{equation}
\left.\frac{\partial^2 V}{\partial \phi^2}\right\vert_{\phi=0}
=\frac{m^2 g^2\Lambda^2}{4\pi^2}\left(\frac{4\pi^2}{g^2\Lambda^2}-1\right) ,
\label{ddV}
\end{equation}
we see that if  $\frac{4\pi^2}{g^2\Lambda^2}>1$ then the extremum at $\phi=0$ corresponds to a minimum, while for $\frac{4\pi^2}{g^2\Lambda^2}<1$
we have a maximum at the origin. Defining $g_c$ as
\begin{equation}
g_c^2\equiv \frac{4\pi^2}{\Lambda^2}
\label{gc}\end{equation}
we see that for a weak coupling $g<g_c$ we have a minimum at the origin, while at strong coupling $g>g_c$ we have a maximum.
The type of extrema at the origin of the potential corresponds to the value of the coupling.\\
Now let us determine the second (possible) extreme of the potential. Since the r.h.s of the second equation in eq.(\ref{dv1}) is negative (i.e. $ x \log\left(\frac{x}{1+x}\right) \leq 0$) this equation has a solution only for a strong coupling $g>g_c$.
A value for $x$ (or that of the scalar field $\phi$), at the minimum can not be solved analytically, since the second equation in eq.(\ref{dv1}) is a transcendental equation.
One way to determine a solution is to seek for the intersection between the curve of the funtion $ x \log\left(\frac{x}{1+x}\right)$ r.h.s. in the second eq. (\ref{dv1}), and the constant in the l.h.s.
In this case do exist an intersection (only one, as the r.h.s. is a monotonic function), giving a solution for the $x$ variable, leading in its turn to a non-trivial solution in $\phi=\phi_{min}$ which is a \textit{minimum}\footnote{According to definition eq. (\ref{xphi}), $x$ is a quadratic function in $\phi$: $x\sim \phi^2$, so for a given value of $x$ we have two solutions in $\phi$ related by a change of sign. Due to this symmetry, we will allow ourselves to refer to only one solution.}
The extremum in this case corresponds to a minimum.
Notice that in all cases we have at  large $x$  the limit  $V \rightarrow \infty$  for $x  \rightarrow \infty $ regardless of the value of the coupling $g$.\\
\begin{figure}[h] \centering \includegraphics[scale=.6]{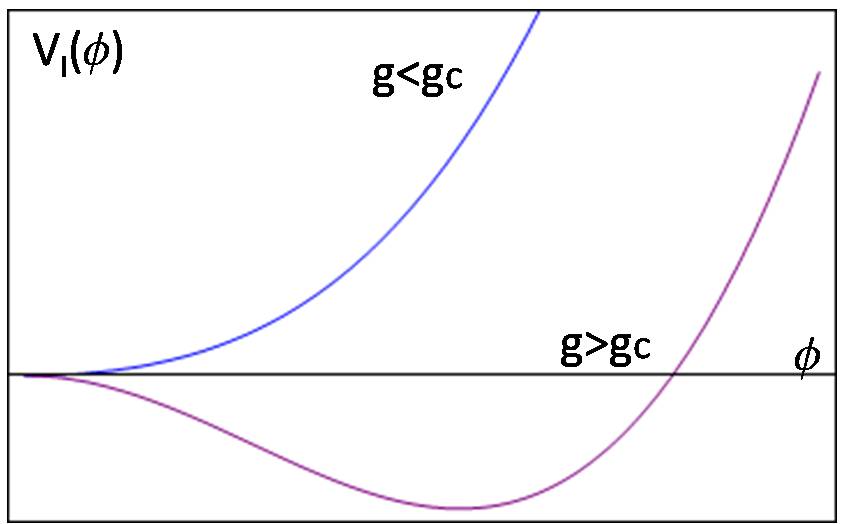}
\caption{Effective potencial (\ref{vefphi}) as a function of $\phi$. The critical value of the coupling $g_c$, separates two kinds of behaviours.} \label{figpotVI} \end{figure}

\noindent Therefore, we have: if $g<g_c$, the potential minimizes in the origin $\phi=0$; whereas for $g>g_c$, the potential minimizes in a non trivial value $\phi=\phi_{min}$ (figure \ref{figpotVI}). The value of the coupling $g=g_c$ define a critical value separating between both behaviours of the potential.\medskip\\
From equation (\ref{phi}) the field $\phi\sim \bar{\psi}\psi$, is a Lorentz invariant quantity, so  $\phi$ is scalar field. When the field $\phi$ is stabilized, a non trivial expectation value reflects the presence of a fermion condensate. \\
Now, if the field has an expectation value $<\phi>=0$, it means that the state of paired fermions $\bar{\psi}\psi$ is not present, so we have a system consisting in the original massless fermion particles with a 4-Fermi interaction, and a condensate is not energetically favoured. This happens for a "weak" coupling $g<g_c$. On the other hand, if the expectation value $<\phi>\neq 0$, then we have a fermion condensate represented effectively by the scalar field. This happens for a "strong" coupling $g>g_c$, and a fermion condensate is dynamically formed since it reduces the energy of the system.\\
Thus, we see that two different fluid phases (massless fermions or fermion condensate) are obtained depending on the strength of the coupling. Next, we investigate the cosmological dynamics of each of these fluids.\\


\section{Standard Cosmology.}\label{cosmo}
The widely accepted current standard cosmological model (the Big Bang theory) is based in Einstein's theory of General Relativity. If conditions of spatial homogeneity and isotropy are assumed, it may be obtained the so called FRWL equations (Friedmann -Robertson -Walker -Lemaitre). As these assumptions agree with observations\footnote{CMBR is a smooth bath of radiation, whereas Large Scale Structure reveal uniform distribution of matter at cosmological scales, with $\gtrsim 100 Mpc$.} to a very high precision, we will use this same theoretical framework. Because the necessary equations are well known and their deduction can be found in standard text books, in the following we limit ourselves to write them and to give only a brief explanation.\\
The equation
\begin{equation}
{\dot{a}^2\over a^2} +{k\over a^2}={8\pi G\over 3}\rho ,\label{frw}
\end{equation}
relates the expansion rate (in time) of the scale factor $a$, and the curvature $k$ of the universe, to the total energy density $\rho$.
 Remember that the curvature parameter is allowed to take the values +1, 0, -1, corresponding to spherical, flat and hiperbolic geometry respectively. Along this paper we will \textit{always} take a flat geometry $k=0$, as suggested on the one hand from the \textit{theory} of early cosmological inflation, and on the other hand (and most important) from \textit{observation} of the CMBR. The Hubble parameter is given $H=\dot a /a$ and we are going to work in a  flat universe ( $k=0$) so eq. (\ref{frw}) reduces to
\begin{equation}
H^2\equiv \left(\frac{\dot a}{a}\right)^2={8\pi G\over 3}\rho .\label{Hrho}
\end{equation}
The continuity equation for a fluid with energy density $\rho$ and pressure $P$ is
\begin{equation}
\dot{\rho}+3H(\rho +P)=0.\label{cont}
\end{equation}
A perfect (barotropic) fluid  satisfies the equation of state $P_\alpha=w_\alpha \rho_\alpha$, with $-1<w_\alpha<1$ a constant, eq. (\ref{cont}) can be solved analytically.  From the cosmological point of view, the substances contained in the universe can be described as radiation, which has $w_r=1/3$, and matter (dust) having $w_m=0$ (besides the Dark Energy component or Baryons contribution). For those we have respectively
\begin{equation}
\rho_r= \rho_{ri}\left({a\over a_i} \right)^{-4},\quad\rho_m= \rho_{mi}\left({a\over a_i} \right)^{-3}. \label{rhos}
\end{equation}
A scalar field $\phi$, with a self-interaction potential $V(\phi)$, has energy density $\rho_\phi$ and pressure $P_\phi$ given by
\begin{equation}
\rho_\phi =E_k +V(\phi),\quad P_\phi =E_k -V(\phi),\quad E_k={1\over 2}\dot{\phi}^2
\end{equation}
where we have also denifed the kinetic energy $E_k$ in the third equation.
Considering an universe containing radiation, matter and a scalar field, the total energy density is written
\begin{equation}
\rho =\rho_r +\rho_m +\rho_\phi .\label{rhotot}
\end{equation}
For a given component fluid "$\alpha$", it is useful to know its relative density, defined as the ratio of its energy density to the total energy density:
\begin{equation}
\Omega_{\alpha}={\rho_{\alpha}\over \rho} ={8\pi\mbox{G} \rho_{\alpha}\over 3H^2}, \label{Omega1}
\end{equation}
where we have used eq. (\ref{Hrho}) in the second equality. In a flat universe one has  the condition
\begin{equation}
\Omega_r +\Omega_m +\Omega_\phi =1.
\label{OmegaT}\end{equation}
It is interesting to note that while eq.(\ref{OmegaT}) remains valid even when we have a negative  $\rho_{\alpha}$, the
quantity  $\Omega_{\alpha}$ is no longer constrainted to the values $0\leq \Omega_{\alpha}\leq 1$. In the work presented here, the fluids can have a negative energy density, giving  $\Omega_{\alpha}<0 $, or a total energy density $\rho$ that vanish at finite values of the scale
factor $a(t)$, in which case we would have $\Omega_{\alpha}  \rightarrow \pm \infty$.\\
Taking the time derivative in equation (\ref{Hrho}) and using eq. (\ref{cont}), it can be found
\begin{equation}
\dot{H}=-{1\over 2}\left(\rho_m +{4\over 3}\rho_r +\dot{\phi}^2\right). \label{Hpunto}
\label{dH}\end{equation}
Note that the r.h.s. in eq.(\ref{dH}) is always negative.
The equation of motion for a spatially homogeneous scalar field, (a modified Klein-Gordon equation) is given by
\begin{equation}
\ddot{\phi}+3H\dot{\phi}+\partial_{\phi}V=0.\label{ecmov}
\end{equation}
It is also useful an equation for the acceleration of the scale factor:
\begin{equation}
\frac{\ddot a}{a}=-\frac{4\pi \mbox{G}}{3}(\rho +3P). \label{ac}
\end{equation}
Differential equations (\ref{Hrho}), (\ref{Hpunto}), (\ref{ecmov}), together with (\ref{rhos}) constitute a complete set which can be solved numerically (since we can not always write an analytical solution). Nevertheless, it is convenient to attempt to outline the general  behaviour
of the dynamical system. Thus, before going to solve for our NJL potential, let us point out the following generic facts:\\
The evolution of the scalar field is such that it will minimize the scalar potential $V(\phi)$, so for an arbitrary initial value $\phi_i$, the field will roll to lower values of the potential, in such a way that eventually it will adopt a constant value ($\phi=\phi_{min}$ being the minimum).
Given than the scale factor is a positive defined quantity, the energy densities for matter and radiation eq. (\ref{rhos}) are always positive quantities and never equal to zero for finite values of the scale factor $a(t)$. So, the total energy density eq. (\ref{rhotot}) remains always positive as long as the condition
\begin{equation}
\rho=\rho_r +\rho_m +\rho_\phi > 0\label{rhopositiva}
\end{equation}
is satisfied. Thus, equation (\ref{Hrho}) says that $H=0$, that is $\dot{a}= 0$, \textit{never} happens (eq. (\ref{Hrho})) as long as $\rho\neq 0$. This implicates that $\dot{a}>0$ \textit{always}. This means that the scale factor $a(t)$ never reaches an extremum value along its time evolution (taking an initial condition $H_i>0$, since we know  that the universe is expanding at present time).\\
Nevertheless, it is interesting o note that there is no known physical principle forbiding the existence of a fluid  with a negative potential $V(\phi)<0$, at least for some values of the field $\phi$. In this case, it could well happen that equation (\ref{rhopositiva}) become an equality, meaning $\rho=0$ for finite values of $a(t)$, which in turn implies $H=0$, and $\dot{a}=0$; i.e, the scale factor reaches an extremum value (indeed a maximum, since as seen before, it was initial growing). Now, equation (\ref{Hpunto}) imposes an always decreasing Hubble parameter $H$ (because the right hand side is always negative), so that after being $H=0$ it must be $H<0$, and therefore $\dot{a}<0$, i.e. the scale factor decrease. In other words, the universe must be contracting after reaching its maximum size. Observe that this result is a consequence only of the negativity of the potential, and it is independent of its specific form. This collapsing universe is valid even for a flat universe $k=0$.
To conclude, if a fermion condensate is energetically favored then the minimum of potential $V(\phi)$ is negative and the universe will recolapse.

\section{Dynamics of Massless Fermions Phase (weak coupling $g<g_c$).}\label{dyn.weak}

As we have seen in section \ref{secnjl}, for a weak coupling $g<g_c$ the minimum of the potential $V(\phi_{min})=0$ is located at the origin with $\phi_{min}=0$, and $V$ does not take negative values. Therefore, the total energy density and $H$ never vanish for finite values of the scale factor $a$, and we have $\dot{a}>0$ due to eq. (\ref{Hrho}). So the scale factor $a(t)$ is always growing, going to an infinite size in an infinite time. Now, from equation (\ref{ac}), it can be seen that, in order to have $\ddot{a}<0$, i.e. the universe to slow down its expansion rate, then
\begin{equation}
{2\over 3}\rho_r +{1\over 2}\rho_m +\dot{\phi}^2 >V(\phi) \label{accond}
\end{equation}
is a condition to be satisfied.
This, of course, in not always the case: we could take an initial field amplitude $\phi_i$ as big to make the initial value of the potential $V_i=V(\phi_i)$ big enough so that inequality (\ref{accond}) does not hold, and we would have instead $V>{2\over 3}\rho_r +{1\over 2}\rho_m +\dot{\phi}^2$. In this case we could have an acceleration of the scale factor, i.e. an accelerating universe, though it would be an "early" acceleration, as it would be present an initial times, i.e. before letting the fluid densities to dilute and field to evolve. As time passes, the field rolls down minimizing the potential, and eventually acquires some value $\phi<\phi_i$ such that condition (\ref{accond}) become fulfilled.
\begin{figure}[h] \centering \includegraphics[scale=.5]{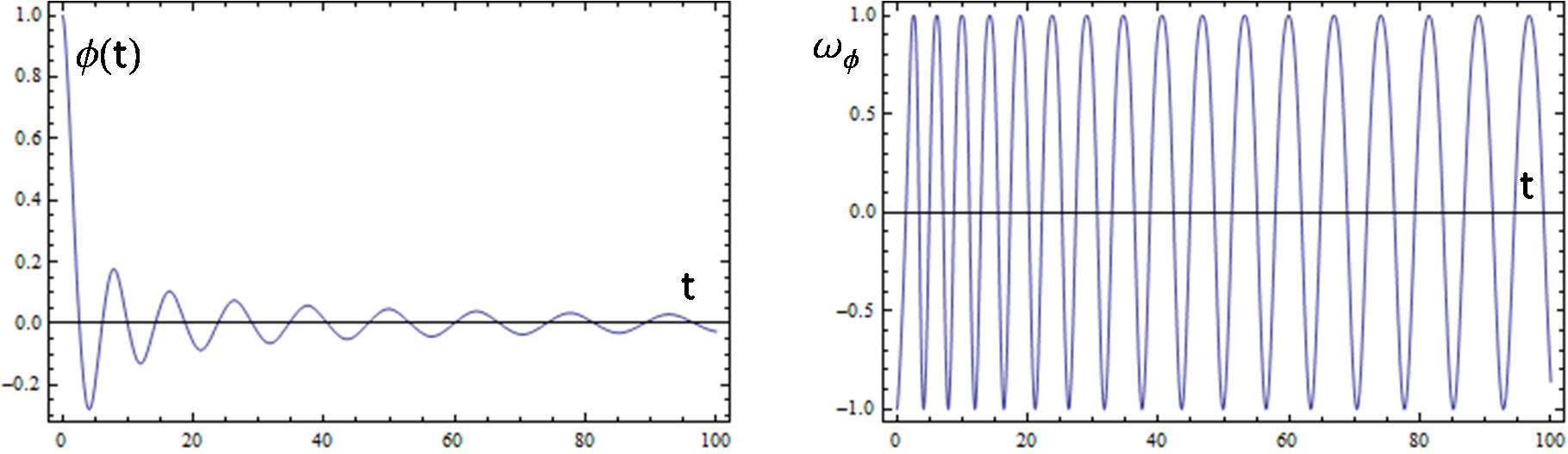}
\caption{\textit{Left:} Scalar field amplitude $\phi$. \textit{Right:} State equation coefficient $\omega_\phi$. Both variables are shown as functions of time.} \label{uncampofig03}
\end{figure}
\begin{figure}[h!] \centering \includegraphics[scale=.5]{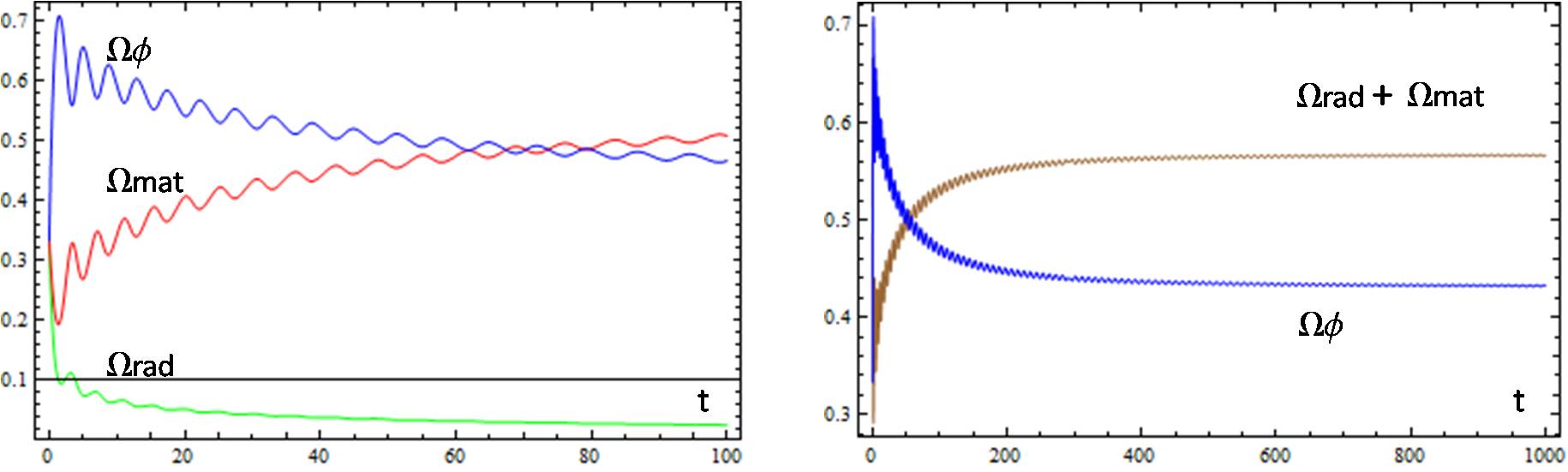}
\caption{\textit{Left:} Relative densities $\Omega_\alpha$ for radiation, matter and field. \textit{Right:} Total relative density for barotropic fluids (matter and radiation) and for the field $\phi$. The horizontal axis in both graphics represents time. Note that we show a different scale of time in each plot for the same solution.}
\label{uncampofig02}
\end{figure}
\begin{figure}[h!!] \centering \includegraphics[scale=.5]{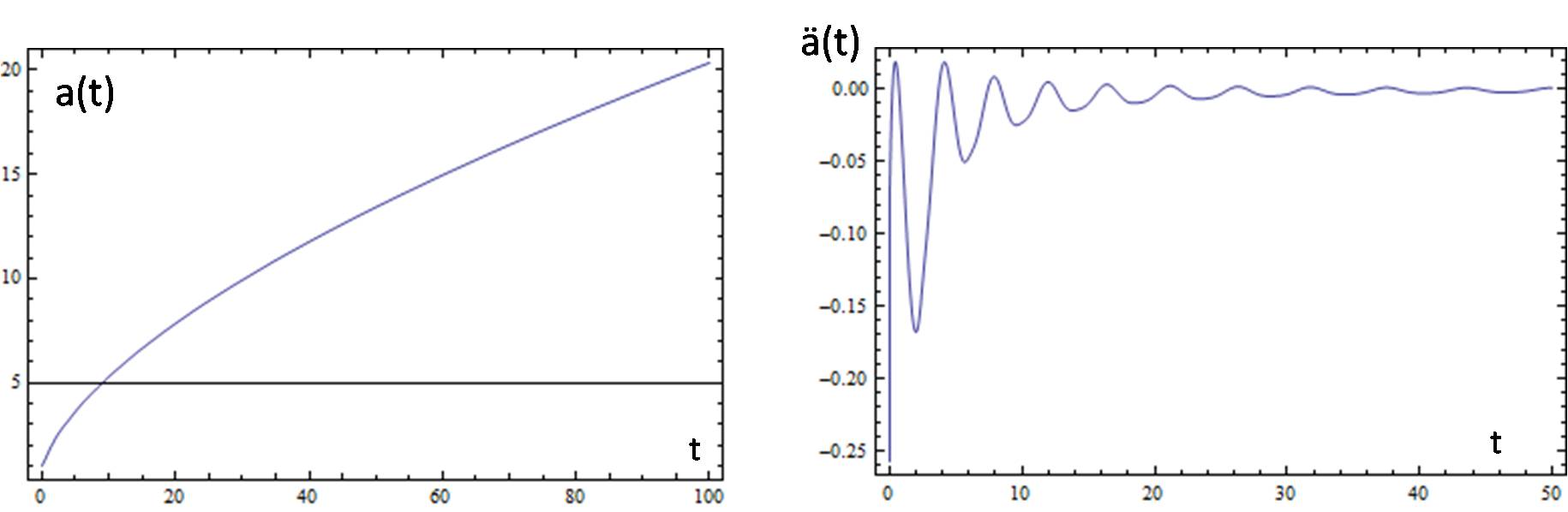}
\caption{\textit{Left}: Scale factor $a(t)$. \textit{Right}: acceleration of the scale factor $\ddot{a}(t)$. Both variables are shown as functions of time. Note that $\ddot{a}(t)$ adopt mostly negative values (tends to zero from below).}
\label{uncampofig01}
\end{figure}
\begin{figure}[h!!] \centering \includegraphics[scale=.6]{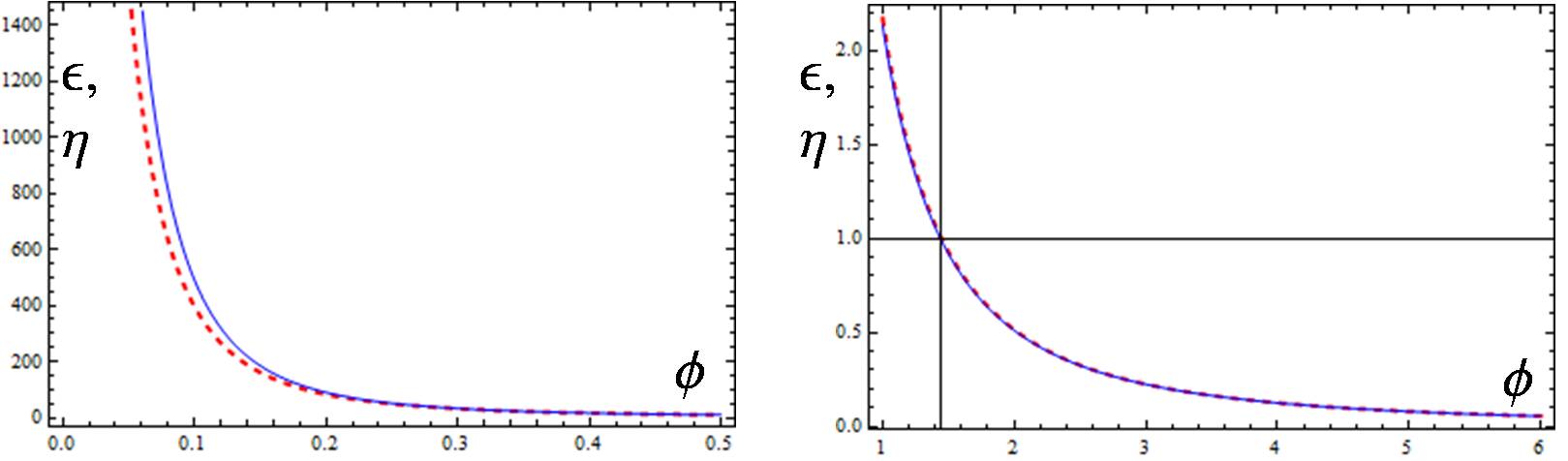}
\caption{Slow Roll parameters $\epsilon$ (dashed-red curve), and $\eta$ (continuous-blue). \textit{Left}: from $\phi=0$ to $\phi=0.5$.  \textit{Right}: from $\phi=1$ to $\phi=6$. Only in the region $\phi\gtrsim 1.4$ aprox. (and further on) one can expect the acceleration conditions $\epsilon<1$, $\eta<1$ to be satisfied.}
\label{uncampofig04}
\end{figure}
Given that the densities of matter and radiation never reach a null value in a finite time, and that the field amplitude tends to be stabilized around the minimum (i.e. $\phi\to 0$), for a big enough amount of time, we expect a vanishing potential and velocity, $V\sim 0$, $\dot{\phi}\sim 0$ to be a good approximation to a final situation, in which (\ref{accond}) is still satisfied.\\
We show an example of numerical solution in the figures. In fig. \ref{uncampofig03} we see that the field has a damped oscillation around $\phi=0$, and in consistency with this, its kinetic energy (velocity) diminish in time. Simultaneously, the potential valuated at $\phi\sim 0$ goes to lower values (according to $V(\phi_{min})=0$). We can see that although the universe is expanding, it always ends up in an non-accelerating regime (figure \ref{uncampofig01}). A Taylor expansion for the potential about $\phi=0$ gives
\begin{equation}
V\simeq {1\over 2}m^2\left( 1- {\Lambda^2 g^2\over 4\pi^2}\right)\phi^2 ,
\end{equation}
where the whole coefficient multiplying on $\phi^2$, is a positive quantity, as $g^2<4\pi^2/\Lambda^2$.
The coefficient of state $\omega_\phi$ defined below eq. (\ref{cont}), for the field $\phi$, written explicitly is
\begin{equation}
\omega_\phi={P_\phi\over \rho_\phi} ={E_k -V\over E_k+V}. \label{omegaphi}
\end{equation}
Since at late times, when the field oscillates around its minimum with a quadratic potential, the average value is $<\omega_\phi>=0$ and $\rho_\phi$ evolves as matter with $\rho_\phi \propto a^{-3}$ [ ref. de la Macorra '99].\\
Within the context of Early Cosmic Inflation theory, the so called \textit{Slow Roll} parameters are defined as follows:
\begin{equation}
\epsilon ={M_p^2 \over 2}\left( {V'\over V}\right)^2,\quad
\eta =M_p^2\left( {V''\over V}\right),
\end{equation}
which have to satisfy the conditions $\epsilon<1$, $|\eta|<1$ in order to the potencial may cause a positive acceleration.
Even though they are valid for a single field, whitout additional fluids (matter and/or radiation), we show them in fig. \ref{uncampofig04} the Slow Roll parameters, for the seek of completeness.\medskip

\section{Fermions Condensate Dynamics (strong coupling, $g>g_c$).}\label{dyn.strong}

\begin{figure}[h!] \centering \includegraphics[scale=.5]{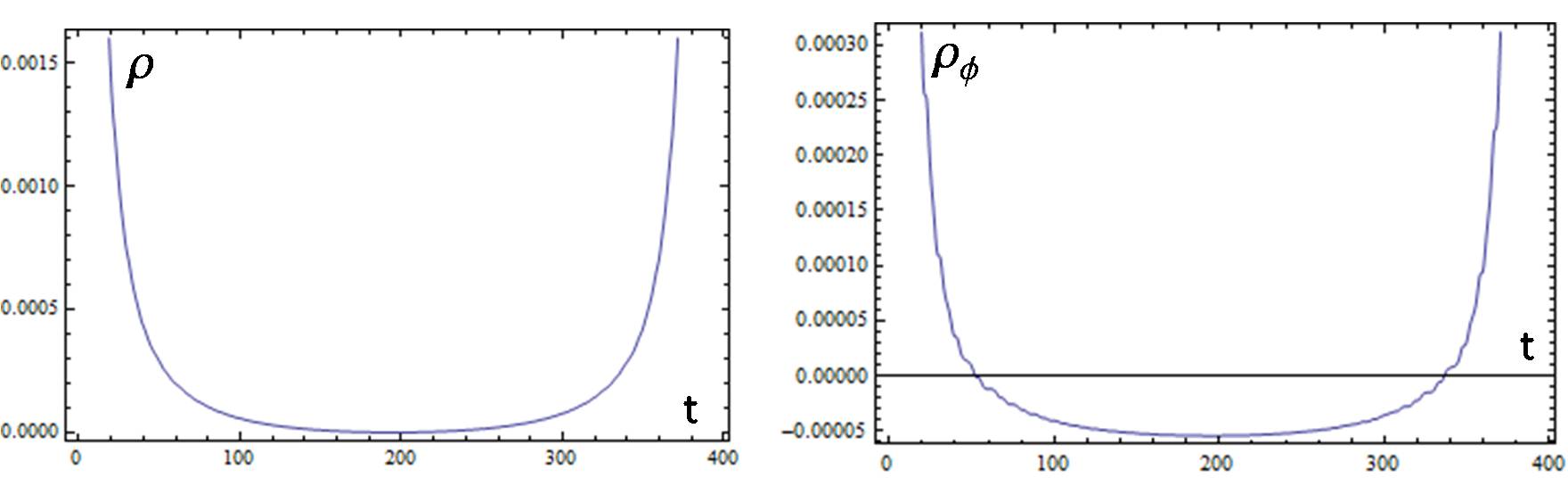}
\caption{\textit{Left:} Total energy density $\rho=\rho_r +\rho_m +\rho_\phi$. It is a positive quantity, but vanishes at a single point, near $t\simeq 200$ aprox. \textit{Right:} Energy density of the field. It is a null quantity ($\rho_\phi=0$) twice: one time in the expansion phase (near $t=60$ aprox.), and again in the contraction phase (about $t=340$ aprox.); and becomes a negative quantity in between.} \label{energydensities2}
\end{figure}

\begin{figure}[h!] \centering \includegraphics[scale=.55]{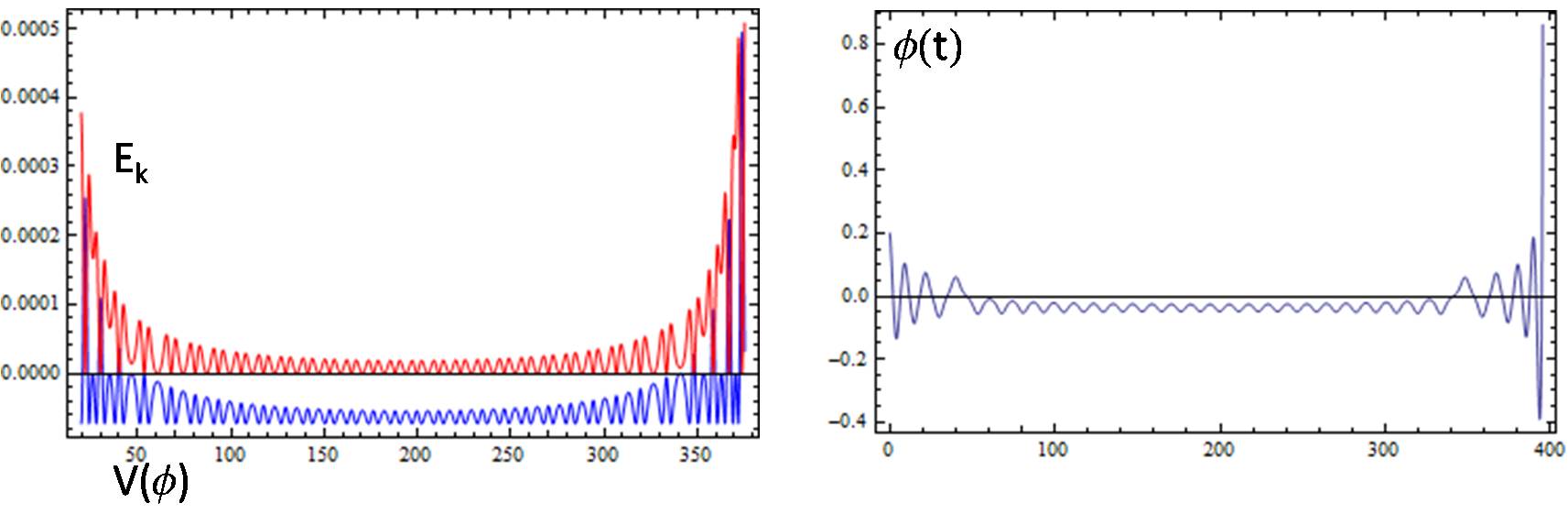}
\caption{\textit{Left}: Although the kinetic energy (red-upper curve) is zero initially, it overtakes the potential energy (blue-lower curve) and remains dominant all the way even to the collapsing time when $a(t)=0$.
\textit{Right}: The field oscillates around $\phi_{min}$ and is becoming divergent as getting close to $t\simeq 400$, which is the time when $a(t)\to 0$. }\label{field2}
\end{figure}

\begin{figure}[h!] \centering \includegraphics[scale=.5]{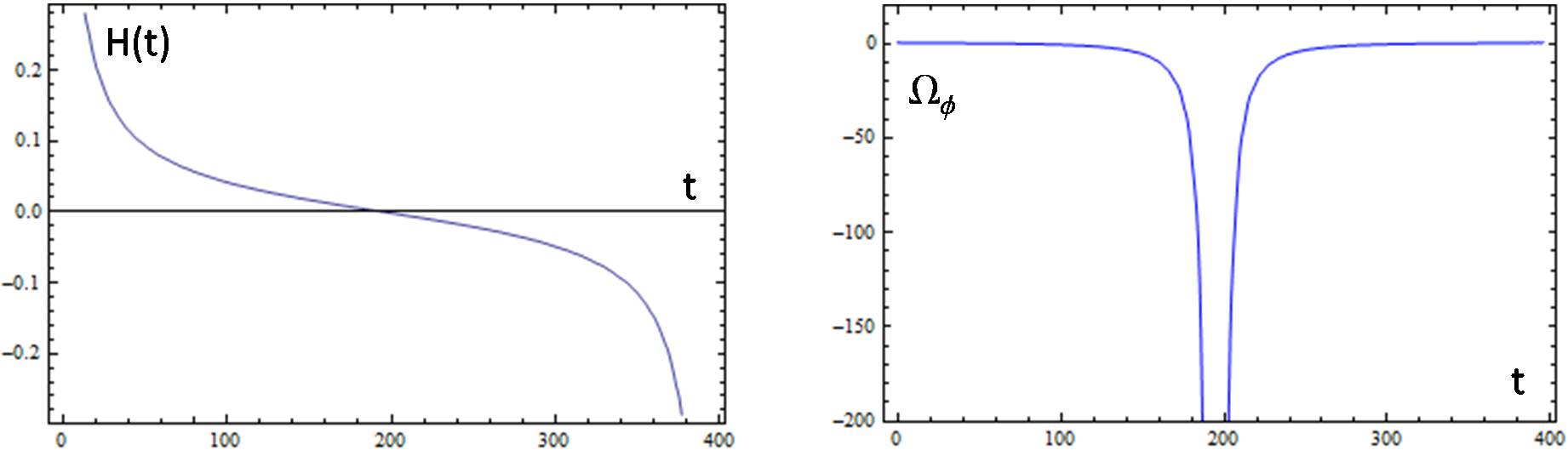}
\caption{\textit{Left:} Hubble parameter. It is a null quantity about $t\simeq 200$ aprox. \textit{Right:} Relative density of the field. As $H(t)$ vanish, $\Omega_\phi$ becomes a divergent quantity near the null point.} \label{Homega2}
\end{figure}

\begin{figure}[h!] \centering \includegraphics[scale=.55]{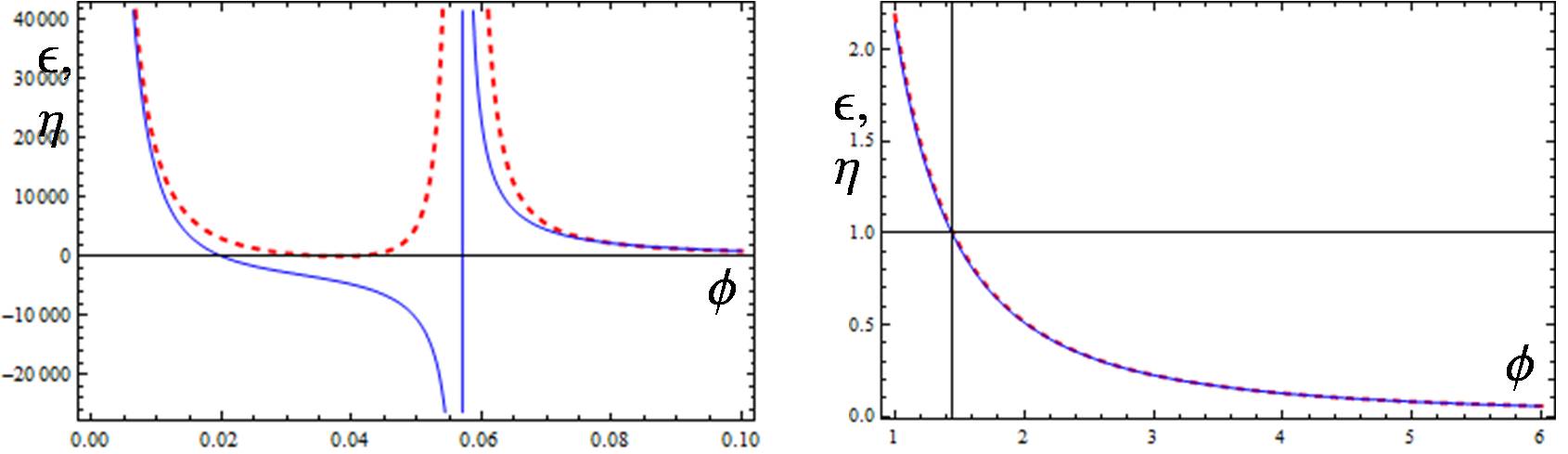}
\caption{Slow Roll parameters $\epsilon$ (dashes-red curve), and $\eta$ (continuous-blue). \textit{Left}: From 0 to 0.1 in $\phi$. \textit{Right}: From 1 to 6 in $\phi$. Only in the region $\phi\gtrsim 1.4$ aprox. (and further on) one can expect the acceleration conditions $\epsilon<<1$, $\eta<<1$ to be satisfied.}\label{slowroll2}
\end{figure}

\begin{figure}[h!] \centering \includegraphics[scale=.5]{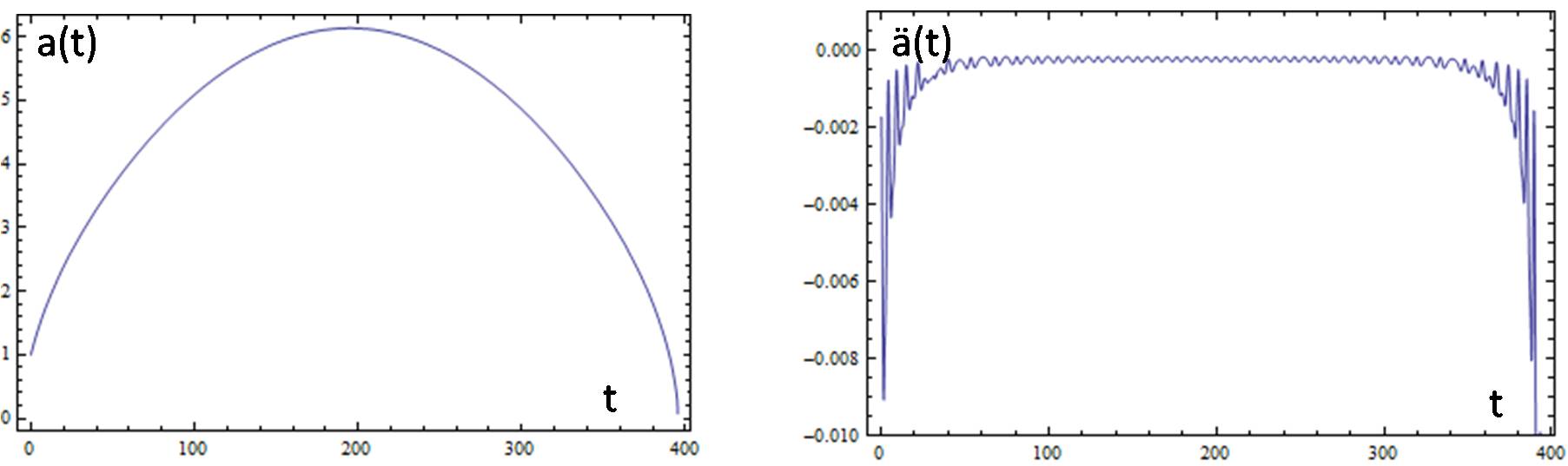}
\caption{\textit{Left}: Scale factor. \textit{Right}: Acceleration $\ddot{a}(t)$. Both plots are to be interpreted as describing a universe which expands without acceleration (note that $\ddot{a}(t)$ is never greater than zero), reaching a maximum value about $t\simeq 200$ approx., thereafter falling in a contracting phase all the way long to collapse.}
\label{aac2}
\end{figure}

The strong coupling case leads to a fermion condensate and therefore to a negative potential $V$ at its minimum. The potential has at the origin $V(\phi=0)=0$ and decreases to negative values for $0<\phi<\phi_{min}$. For  $\phi>\phi_{min}$ it grows monotonically, eventually passing from negative to positive values.
Let us consider at first the simpler approach of a universe containing only a scalar field ($\rho_r=\rho_m=0$, i.e. no additional fluids), evolving under a \textit{generic} potential possessing a negative value $V_{min}=V(\phi_{min})<0$ when minimized.
If the initial velocity $\dot{\phi}_i=0$, then the kinetic energy of the field has a null value as well, so we have for the initial energy density ${\rho_\phi}_i=V_i$. The initial amplitude for the field $\phi_i$ can not be such that makes $V(\phi_i)<0$, because it would lead to an imaginary value for $H$, according to eq. (\ref{Hrho}). Thus, we must take always $\phi_i$ such that $V_i>0$. As before we begin with $H_i=1>0$, therefore equation (\ref{Hrho}) says that $a(t)$ initially is increasing in time. The equation (\ref{Hpunto}) is written $\dot{H}=-(1/2)\dot{\phi}^2$, so that $H$ \textit{always diminish} in time. As the potential is minimized, it goes from positive to negative values, and from equation (\ref{Hrho}) eventually it will be $H=0$, and after this $H<0$, corresponding respectively to $\dot{a}=0$ and $\dot{a}<0$. In words this means that after an initial period of expansion (increasing scale factor), a maximum value is reached, followed by a period of contraction.
Since $\dot{H}$ remains always negative, then $a(t)$ will continue decreasing, so that it necessarily will collapse. In other words, it will be $a=0$ in a \textit{finite} time in the \textit{future} (because the evolution is forward in time: the field minimizes, not otherwise).\\
Now, while the expanding phase is taking place, the field is rolling down, eventually entering in a damped oscillatory regime nearly the minimum, where the potential has become negative, $V_{min}<0$. Because of the damping, the kinetic energy tends to a zero value, $E_k\to 0$. Thus, the energy density of the field $\rho_\phi=E_k+V$ goes from positive values (near $\phi_i$) to negative values (near $\phi_{min}$), so at some time in between, it is $\rho_\phi=0$.
The total energy density, as well as the individual densities for each fluid (if there were additional fluids), would go to \textit{diminish} in time (as can be seen for radiation and matter in eqs. (\ref{rhos}) with $\rho_\alpha\sim a^{-n}$, and $a(t)$ increasing). By a similar reasoning, because $a(t)$ is decreasing in the \textit{contracting} phase, the energy densities behave the opposite way, i.e. they all \textit{increase} in time. Therefore, we expect $\rho_\phi=0$ to happen twice. In its turn, this implicate that the coefficient of state $\omega_\phi$, eq. (\ref{omegaphi}) become a divergent quantity also twice, around this two points, and near them, $\omega_\phi$ is not anymore a useful parameter to characterize the fluid represented by the field $\phi$. Below we show a numerical solution example (figs. \ref{energydensities2}-\ref{aac2}).\\
\noindent As we mention before, in section \ref{cosmo}, a similar circumstance arises in dealing with the relative densities $\Omega_\alpha$: it is considered that in order to this parameter to make sense, a relative density should adopt values $0\leq \Omega_\alpha \leq 1$. However, as can be seen in eq. (\ref{Omega1}), if at some time is $H=0$, then nearly this value, each $\Omega_\alpha$ turns into a divergent variable. The situation is even weirdest for the field, because near the minimum it is $\rho_\phi\sim V_{min}<0$, the energy density of the field is similar to the potential, which is negative. This would make $\Omega_\alpha\to -\infty$ (a divergent and \textit{negative} relative density!).\\

\noindent Consider now a universe containing matter and radiation in addition to our NJL fluid. An interesting question is, may the presence of these fluids prevent the universe to collapse?
Remember that the condition for an increasing scale factor can be reduced to the inequality (\ref{rhopositiva}). If the scale factor is supposed to grow forever, this condition must be hold \textit{always}. Now, according to the explanations given above, initially the scale factor is growing indeed. Thus, from eq. (\ref{rhos}) we see that the densities of both barotropic fluids (matter and radiation) must be decreasing. At the same time, because the field is stabilizing in the minimum of the potential, the kinetic energy of the field $E_k=(1/2)\dot{\phi}^2$ is diminishing to zero, whereas the potential is going to a constant value $V\to V_{min}$, in such a way that necessarily, condition (\ref{rhopositiva}) ceases to hold. Therefore, even in presence of additional barotropic fluids (does not matter the relative amount with respect to that of the fluid associated with the field), the collapsing universe situation can not be avoided.\\

\noindent The previous qualitative generic analysis is verified by the numerical solution for our NJL potential in particular (figures \ref{energydensities2}-\ref{aac2}). By observing the graphics, we found an unpredicted interesting non-trivial behaviour of the field amplitude: while the scale factor undergoes the expanding, and contracting phases successively, an damped oscillating phase around $\phi_{min}$ is taking place, as expected. But then, at some point in the contracting phase, the field amplitude goes to bigger values, and as the scale factor approach to $a=0$, the field is taken out from the minimum and it begins to increase monotonically!\footnote{It could decrease instead, depending on the initial conditions. Whatever the case, the monotonic growing in absolute value is an unexpected behaviour, which do happen indeed.} Is this an acceptable result? Intuitively, as $a$ is decreasing, it is reasonable to expect all densities to be growing. In particular, if the field density $\rho_\phi=E_k+V$ is getting bigger, it should be due to an increase in the field velocity (so $E_k$ gets bigger), or in the field amplitude (so $V$ gets bigger); or both. This behaviour can indeed be explained observing eq. (\ref{cont}).
The energy evolution of a barotropic fluid $\rho_b$ is given by
\begin{equation}
\dot{\rho_b}=- 3H(\rho_b +P_b)= - 3H\rho_b (1+w),
\label{ax1}\end{equation}
and for a scalar field with energy density $\rho_\p = E_k + V(\phi)$ and pressure  $ P_\p= E_k - V(\phi)$ and $E_k=\dot\phi^2/2$
\begin{equation}
\dot{\rp}=- 3H(\rp +P)= - 3H \dot\phi^2= - 6 H E_k .
\label{ax2}\end{equation}
We can see from  eqs.(\ref{ax1}) and (\ref{ax2}) that for  a positive barotropic fluid $\rho_b $ with an EQS $ |w|<1$, the sign of $\dot\rho_b$  and  $\dot\rp$  are   negative as long as $H$ is positive while they become negative for $H<0$. Therefore $\rho_b,\;\rp$ are decreasing functions as a function of time for  $H>0$ and increasing  for $H<0$. Since we have seen that $\dot H$ is negative, this implies that H is always a decreasing function of time. If $H$ can vanish at a finite time only if $\rp$ becomes negative, i.e.  if the potential $V$ becomes negative and  $E_k=-V$ at say $t=t_c$. After this time $H(t>t_c)$ becomes negative and will remain negative for $t>t_c$ and  $\rho_b $ and $\rp$ will start growing  with time for $t\geq t_c$.\\
\noindent Figure \ref{field2} show both kinetic and potential energies, and we can see that even though the inicial kinetic energy is zero, it overtakes the potential energy and remains so until the collapsing moment $t_{final}$ when $a(t)=0$ at late times. Nevertheless, the potential energy also grows as the time is approaching $t_{final}$, so the field amplitude is eventually expeled from oscillating about the minimum.\medskip

\section{NJL fluid with a Cosmological Constant.}\label{coscte}

Due to its theoretical properties and observational requirements, a Cosmological Constant is a very usual and useful ingredient included in cosmological models, and it is worth to consider such contribution in our model. Its defining property is an energy density $\rho_\Lambda$ which does not vary in time, and a coefficient of state $\omega_\Lambda=-1$, which gives a pressure $P_\Lambda=-\rho_\Lambda$. In a universe containing \textit{only} a Cosmological Constant, the equation (\ref{ac}) is written $\ddot{a}=(8\pi G/3)a\times\rho_\Lambda$ which, as $\rho_\Lambda>0$, implicates $\ddot{a}(t)>0$ \textit{always}. Therefore, such an universe is always accelerating its expansion. In fact, in this case the equation (\ref{Hrho}) may be solved analytically, after substituting eq. (\ref{Hrho}), giving the well known solution $a(t)=a_i\exp (t\sqrt{8\pi G \rho_\Lambda /3})$.
How do the presence of a Cosmological Constant affect our previous considerations of a universe including our NJL fluid, besides matter and radiation components?
Will the universe accelerate or collapse, even in the presence of a scalar field with a negative potential $V<0$?
Because the density $\rho_\Lambda$ is constant, we have that the differential equations are not modified, other than just adding a term in the expression for $H$, equation (\ref{Hrho}). In particular, the equation of motion eq. (\ref{ecmov}) remains unchanged, so the field dynamics is not affected. As before, we have to deal with two cases.\\[3mm]
\textbf{a) Free Fermions ($g<g_c$).} As studied before, the potential is $V\geq 0$, and its minimum value is $V_{min}=0$. Also, with the pass of time, both matter and radiation densities dilute, going to vanish. From equation (\ref{ac}), it can be deduced the condition for universe to decelerate:
\begin{equation}
\rho_\Lambda <\rho_r +{1\over 2}\rho_m +2E_k -V(\phi)\quad \mbox{(for $\ddot{a}<0$)}. \label{accond2}
\end{equation}
Given that the left hand side in this inequality is diminishing in time, whereas the right hand side remains constant, we have that eventually this inequality can not hold anymore, and becomes an equality, meaning $\ddot{a}=0$. This points the beginning of the acceleration period, i.e. $\ddot{a}>0$, where the inequality (\ref{accond2}) gets inverted.
Had the initial conditions been such that inequality (\ref{accond2}) were the opposite, then there would be always an acceleration holding always, because the LHS would never go back to grow.\\
Thus, we see that for a free fermions NJL fluid with a Cosmological Constant, the universe necessarily accelerate, the precise moment depending on the amount of energy densities $\rho_m$, $\rho_r$, with respect to that of $\rho_\Lambda$. This can be specified in the initial conditions, which in their turn can be chosen to solve for a realistic model fitting the observations.\\[3mm]
\textbf{b) Fermion Condensate ($g>g_c$).} We found before that, for a strong coupling, the potential is negative when minimized, $V_{min}<0$. Do the universe necessarily accelerate also in this case? In order for this to happen, condition (\ref{accond2}) eventually must turn into an equality, meaning $\ddot{a}=0$. This is a minimal condition to be satisfied, because it points at least the beginning of an acceleration; it remains to be sure that acceleration will be sustained.
Let us label all quantities with a subindex "ac" at time $t_{ac}$, when $\ddot{a}=0$ (vgr. $V(t_{ac})=V_{ac}$). From eq. (\ref{accond}), we have\footnote{$\rho_\Lambda$ does not need a label because it is a constant.}
\begin{equation}
\rho_\Lambda \geq\rho_{rac} +{1\over 2}\rho_{mac} +2E_{kac} -V_{ac}\quad \mbox{(for $\ddot{a}\geq0$)}. \label{accond3}
\end{equation}
Remember that the potential take positive values as well as negative ones, so both possibilities must be taken into account. Certainly one can find such set of values of $V$ for a given $\rho_\Lambda$ to satisfy the inequality. However, if we rather want to consider realistic models, we should consider plausible values from observations (besides, we would not like to complicate our lives by considering unrealistic generic situations).\\
From definitions (\ref{Omega1}) it can be found that $\Omega_r/\Omega_m=(1+z)r$, where $z$ is the redshift, and $r=\Omega_{r0}/\Omega_{m0}$ says the amount of radiation with respect to that of matter. The subindex "0" refers to current values, i.e. quantities measured "today". Now, the estimate for $z$ (the time when acceleration begins) is around $z\sim 1$; and it has been measured $r\sim 10^{-4}$ (for the seek of simplicity, here we are interested only in orders of magnitude). Then we have $\Omega_{rac}\sim 10^{-4}\times \Omega_{mac}<<\Omega_{mac}$, or $\rho_{rac}<<\rho_{mac}$.
Now, remember that a decelerating period dominated by matter is supposed to have taken place before $\ddot{a}=0$. In order for this to happen, condition (\ref{accond2}) should have to be true before cond. (\ref{accond3}). For $z\sim 1$ (it could be even as big as, lets say $z\sim 10$, as this would not change the essence of the argument) and using cond. (\ref{accond2}) we would have
\begin{equation}
\rho_m >2(\rho_\Lambda +V -2E_k) \quad\mbox{(in order to be $\ddot{a}<0$)}. \label{accond4}
\end{equation}

\noindent If a positive acceleration eventually come up, the above expression is expected to become an equality.
Now, suppose $V>0$. Then, unless $E_k$ decrease even fast, the RHS in the inequality should be decreasing as time passes, because the potential is minimizing. But $E_k$ can not behave like that indeed, as the field is under a damped rolling, not to mention that $E_k$ is never a negative quantity, so the sum of terms $V-2E_k$ will end up decreasing (would the values of these terms been such that the equality \textit{somehow} would be accomplished at some time, in this case the acceleration \textit{could not be attached} to $\rho_\Lambda$ anyway). On the other hand, for $V<0$, the inequality would become even more strong in time, because again, the potential is minimizing: $V\to V_{min}$, and $0>V>  V_{min}$.
Therefore, if initially the inequality (\ref{accond4}) begins being satisfied, it will remain being so always; in other words, the universe will never accelerate.\\[2mm]

\noindent What about a collapse \textit{in the future}? May the presence of a cosmological constant prevent a decreasing scale factor (time going forward)?
For a growing scale factor we have $\dot{a}>0$, which is true indeed because we take $H_i>0$ is the initial value of $H$.\footnote{Observe this initial condition must be taken to be positive, because otherwise, the universe would be \textit{already} contracting.}
As we explained before, if the scale factor is to reach a maximum $a=a_{max}$, it must be $\dot{a}=0$.
Let us name $t_{am}$ the time when this is accomplished (if so), and label with a subindex "am" the variables valuated at this time.
We have for the total energy density $\rho_{am}=0$, thus $\rho_\Lambda +\rho_{ram} +\rho_{mam} +E_{kam} +V_{am}=0.$
The only way in which this could happen is for $V_{am}<0$. In that case $V_{am}=-|V_{am}|$, so the equation, as a condition to be satisfied by $\rho_\Lambda$, can be written in the more intelligible form
\begin{equation}
\rho_\Lambda =|V_{am}| -E_{kam} -\rho_{mam} -\rho_{ram}\quad\mbox{(to get $\dot{a}=0$)}.\label{amax1}
\end{equation}
If we want to keep our analysis as simple as possible, we may ignore the contribution from radiation, $\rho_{ram}=0$ (observe that, had an acceleration would be possible, then we should assume $t_{ac}<t_{am}$, i.e. acceleration before receding, otherwise the model would not be useful. So, if $\rho_{rac}<<1$ the approximation $\rho_{ram}\sim 0$ is even better, as $\rho_{ram}<\rho_{rac}$).\\
Now, nothing forbids to exist a potential sufficiently deep $V_{min}<0$, so that the equality (\ref{amax1}) can be accomplished.
The exact time at which this is achieved will depend on the relative amounts $E_{kam}$, $\rho_{mam}$, with respect to $\rho_\Lambda$, i.e. on the initial conditions. However, we can estimate a limit value by making $\rho_{mam}\to 0$, $E_{kam}\to 0$, and a stabilized potential $V\to V_{min}$. Then we have
\begin{equation}
\rho_\Lambda =|V_{min}|\quad \mbox{(Maximum allowed value for the universe to collapse)}.
\end{equation}
After $\dot{a}=0$, i.e. $H=0$ (eq. (\ref{Hrho})), the universe must enter into a contraction phase because $H$ is \textit{always} decreasing (eq. (\ref{Hpunto})), meaning $H_{am}\to H<0$, i.e. $\dot{a}<0$. So, eventually the universe will collapse in the future in a finite lapse of time.
For $\rho_\Lambda >|V_{min}|$, the scale factor would never go to contract, as in this case the total energy density $\rho$ would never vanish.\medskip\\
\noindent It is interesting to observe that a Cosmological Constant may be seen as a particular case of a scalar field evolving under a potential stabilized with a positive minimum.
As we have seen, the NJL model has two different behaviours depending on the value of the coupling constant $g$. For weak coupling $g<g_c$ the potential $V(\p)$ has a minimum at the origin with $V(\p=0)=0$ and $V(\p)\geq 0$ otherwise. On the other hand, at strong coupling $g>g_c$ one has a negative minimum $V(\p)|_{min}<0$. So let us approximate the potential $V$ around the minimum and take the ansatz
\be
V(\p(t))=V_o +\frac{1}{2} m^2(\phi(t)-\phi_o)^2,
\label{eqv}\ee
with $V_o$ a constant value (it would be $V_o=0$ at weak coupling and $V_o<0$ at strong coupling) and $\phi_o$ a constant. We can now ask ourself if we can have an accelerating universe. The evolution of the scalar field is just $\ddot\phi ' + 3H\dot\p ' + m^2 \phi'=0$, with $\p '\equiv \phi-\phi_o$
and we could redefine  $\rho_\Lambda+\rho_\p=\rho_\Lambda+E_k +V = \rho_\Lambda + V_o + E_k + \frac{1}{2} m^2(\phi-\phi_o)^2 = \rho_\Lambda + V_o+ E_k+\frac{1}{2} m^2 \phi '^2$ which corresponds to  a massive scalar field with energy density $\rho_\phi'=E_k+\frac{1}{2} m^2 \phi '^2$ in the presence of a cosmological constant $ \rho'_\Lambda= \rho_\Lambda +V_o$. A massive scalar field may accelerate the universe only at large values of $\phi'$ (larger than the Planck mass) when the Slow Roll parameters $\epsilon$ and $\eta$ are smaller than one, while at a late time when the scalar field oscillates around the minimum the energy density $\rho_\phi'$ redshifts as matter, i.e $\rho_\phi'\propto 1/a^3$. In order to have $\ddot a>0$ we must have the quantity $\xi\equiv \rho+3p<0$. So for  a scalar field (with potential  given in eq.(\ref{eqv}))  a barotropic fluid, which we now take for simplicity as matter (without lose of generality),  and a cosmological constant $\rho_\Lambda$, we have $\xi=\rho_m + 4 E_k-2(\rho_\Lambda+V)$. Since the potential $V_o$  vanishes at weak coupling and is negative at strong coupling, there is a cancelation between the two cosmological constants $\rho_\Lambda $ and $V_o$, and the NJL model plays therefore against an accelerating phase around the minimum of the potential, since $V_o$ is negative.\\


\section{Summary of Results and Discussion.}\label{conclu}

The fermion model of Nambu and Jona-Lasinio (NJL) includes two different fermion states resulting from quantum effects, each one being associated with two different physical phases. For a weak coupling $g<g_c$ we have \textit{massless fermion} fluid, whereas for a strong coupling $g>g_c$ a \textit{massive fermion condensate} fluid is obtained. In this later case we can determine the mass of fermions and it is due to non-perturbative effects due to the strong coupling. A very convenient way to describe the system is to consider an equivalent scalar field $\phi$ moving under an effective potential $V=V^\phi_0 +V^\phi_1$, which has a different form depending on the coupling strength. Here we studied the potential and solved the cosmological evolution for each fluid in presence of additional barotropic fluids (e.g. matter-dust or radiation).\\

\noindent For a weak coupling, we found a coefficient of state $\omega_\phi$ with oscillating values around zero, in such a way that the average value $<\omega_\phi>=0$. Also, because the potential goes as $V\sim \phi^2$ near the minimum, we have that the NJL fluid in the form of free fermions dilute as matter. A universe containing such a fluid (with or without matter and/or radiation) will expand forever without accelerating. On the other hand, a universe containing this NJL fluid besides a cosmological constant (with or without matter and/or radiation), will eventually accelerate necessarily, expanding forever.\\

\noindent On the other hand, the strong coupling case (without a cosmological constant) always cause an eventually vanishing energy density. This is due to the fact that the potential is negative when minimized, and even the additional presence of matter and/or radiation do not prevent this to happen. Since the vanishing energy (which is associated with the scale factor reaching a maximum), is followed by a contracting period, this means that a fermion condensate always make the universe to collapse. The energy density of the field $\rho_\phi$ annulates a couple of times (one in the expanding phase, and the another one in the contracting phase). Because of this, some quantities ($\Omega_\phi$, $\omega_\phi$) become inadequate to describe the fluid. It is important to point out the following interesting fact:\\
Equation (\ref{frw}) has been known and well studied since long time ago. If the curvature parameter is $k=+1$, the universe is said to have a \textit{spherical} geometry; the scale factor is expected to get a null value eventually, so we have a collapsing universe. Because a spherical universe is also finite or \textit{closed}, a collapsing universe was always associated with a closed universe. On the other hand, if $k=0$, the universe have a \textit{flat} geometry. For ordinary matter the total energy density could be diminishing, but it could never vanish effectively in a finite time, so the scale factor in this case is expected to be always increasing. Because divergent geodesic lines in a plane never meet again, a flat universe is said to be open. So, an open universe was thought to be infinite in size (although not necessarily, but in any case, always \textit{growing}).
Now, remember that from the beginning, in our present study, we have taken the curvature parameter to be $k=0$, so we have been treating with a flat universe all the time. Nevertheless, we found that, if the universe contains a scalar field with a negative potential, then a future collapse can not be avoided, giving a \textit{collapsing flat} universe! In particular, because a negative potential arises naturally for the NJL model, a collapsing flat universe is also a \textit{natural consequence}.\\

\noindent We also studied a variant of the strong coupling model, consisting in the addition of a cosmological constant. We found that, if the energy density $\rho_\Lambda$ is not big enough to overtake at least the minimized potential $V_{min}$, the eventual receding of the scale factor can not be avoided, and the universe will collapse inevitably. But if $\rho_\Lambda$ exceeds $V_{min}$, then the scale factor will accelerate eventually, and the collapse will be absent.\\

\noindent Perhaps it is worth to emphasize that, in both cases of weak and strong coupling and \textit{without} considering a cosmological constant, one may induce an acceleration of the scale factor by manipulating the initial condition for the field amplitude $\phi_i$, but we do not interest in it because 1) it has to be fine-tuned, and 2) it does not allow to include realistic models in which a previous deceleration period of matter dominance took place.\\
It is important to keep in mind that, once we settle a coupling strength (weak or strong), there is nothing in the theory to allow to switch between them, so actually a phase \textit{transition} can not be considered.\\
A very appealing feature of the NJL model is, in our opinion, the fact that 1) it is based on a "fundamental" symmetry (chiral symmetry), 2) the model leads to a potential which, due to quantum corrections, can adopt negative values in a natural way, and 3) it includes only one parameter: the coupling constant $g$ (two parameters if we count the cut-off $\Lambda$). In return we obtain interesting consequences, as allowing more than one physical phase (each having different cosmological implications), and the possibility of a collapsing universe. This is to be compared with other models involving a symmetry breaking\footnote{For instance in Higgs-like models are required two parameters "$m$" and "$\lambda$" in order to get a potential $V={1\over 2}m^2\varphi^2+{1\over 4}\lambda\varphi^4$, which have to have a "correct" relation between them in order to break the symmetry.} or introducing new kinds of fluids aimed to be relevant to cosmological problems, but at the expense of introducing several fields or parameters.\footnote{For instance, to "justify" the existence of scalar fields with useful potentials, frequently one has to invoke more sophisticated theories, like String, Kaluza-Klein, GUT's, etc. which demand a bigger effort to derive relevant results, and often implicate new exotic physics.}\\

\noindent The issue of how to include a natural phase transition (if possible), for which we consider to modify or "extend" the NJL theory; and whether a positive acceleration may be induced whithout introducing an "artificial" Cosmological Constant, as well as some other interesting questions are left for future publications.


\section*{Acknowledgment}
A.M.  acknowledges financial support from  UNAM PAPIIT Project No. IN101415

\thebibliography{}

\footnotesize{

\bibitem{CMB} A. D. Sakharov.  1966.  Sov.Phys.JETP,22,241,
Primeval adiabatic perturbation in an expanding universe - Peebles, P.J.E. et al. Astrophys.J. 162 (1970) 815-836,
Small scale fluctuations of relic radiation - Sunyaev, R.A. et al. Astrophys.Space Sci. 7 (1970) 3-19,
Probing dark energy using baryonic oscillations in the galaxy power spectrum as a cosmological ruler - Blake, Chris et al. Astrophys.J. 594 (2003) 665-673 astro-ph/0301632

\bibitem{WMAP} Nine-Year Wilkinson Microwave Anisotropy Probe (WMAP) Observations: Final Maps and Results - WMAP Collaboration (Bennett, C.L. et al.) Astrophys.J.Suppl. 208 (2013) 20 arXiv:1212.5225 [astro-ph.CO]

\bibitem{PLANCK}
  P.~A.~R.~Ade {\it et al.} [Planck Collaboration],
  arXiv:1502.01589 [astro-ph.CO],
  P.~A.~R.~Ade {\it et al.} [Planck Collaboration],
  arXiv:1502.01590 [astro-ph.CO].

\bibitem{SN.1}
Perlmutter et al., Astrophys. J. 517, 565 (1999); A. G.
Riess et al., Astron. J. 116, 1009 (1998); R. Amanullah et al., Astrophys. J. 716, 712 (2010),
BV RI light curves for 22 type Ia supernovae - Riess, Adam G. et al. Astron.J. 117 (1999) 707-724 astro-ph/9810291,
\bibitem{SN.2} The Carnegie Supernova Project: First Photometry Data Release of Low-Redshift Type Ia Supernovae - Contreras, Carlos et al. Astron.J. 139 (2010) 519-539 arXiv:0910.3330,
CfA3: 185 Type Ia Supernova Light Curves from the CfA - Hicken, Malcolm et al. Astrophys.J. 700 (2009) 331-357 arXiv:0901.4787,
Ubvri light curves of 44 type ia supernovae - Jha, Saurabh et al. Astron.J. 131 (2006) 527-554 astro-ph/0509234,
New Hubble Space Telescope Discoveries of Type Ia Supernovae at z>=1: Narrowing Constraints on the Early Behavior of Dark Energy - Riess, Adam G. et al. Astrophys.J. 659 (2007) 98-121 astro-ph/0611572 46455850950,
The Hubble Space Telescope Cluster Supernova Survey: V. Improving the Dark Energy Constraints Above z>1 and Building an Early-Type-Hosted Supernova Sample - Suzuki, N. et al. Astrophys.J. 746 (2012) 85 arXiv:1105.3470,

\bibitem{LSS.1}
B. A. Reid et al., Mon. Not. R. Astron. Soc. 404, 60
(2010);  W. J. Percival et al., Mon. Not. R. Astron. Soc. 327, 1297
(2001); M.Tegmark \textit{et al.}
 2DFGRS Collaboration (Colless, Matthew et al.) Mon.Not.Roy.Astron.Soc. 328 (2001) 1039 ,
 SDSS Collaboration (York, Donald G. et al.) Astron.J. 120 (2000) 1579-1587

\bibitem{LSS.2}
Padmanabhan, Nikhil et al. Mon.Not.Roy.Astron.Soc. 427 (2012) 3, 2132-2145 arXiv:1202.0090,
Michael J. et al. Mon.Not.Roy.Astron.Soc. 401 (2010) 1429-1452,
Jones, D.Heath et al. Mon.Not.Roy.Astron.Soc. 399 (2009) 683 arXiv:0903.5451,
SDSS-III - BOSS Collaboration (Dawson, Kyle S. et al.) Astron.J. 145 (2013) 10 arXiv:1208.0022,
BOSS Collaboration (Anderson, Lauren et al.) Mon.Not.Roy.Astron.Soc. 441 (2014) 24-62 arXiv:1312.4877,
BOSS Collaboration (Delubac, Timothee et al.) Astron.Astrophys. 574 (2015) A59 arXiv:1404.1801

\bibitem{BAO.1} SDSS Collaboration (Eisenstein, Daniel J. et al.) Astrophys.J. 633 (2005) 560-574
\bibitem{BAO.2} BOSS DR11 - Tojeiro, Rita et al. Mon.Not.Roy.Astron.Soc. 440 (2014) 2222 arXiv:1401.1768,
 SDSS DR10Isabelle et al. Astron.Astrophys. 563 (2014) A54 arXiv:1311.4870,


\bibitem{NJL}
Y.~Nambu and G.~Jona-Lasinio,
  Phys.\ Rev.\  {\bf 122}, 345 (1961),
 Y.~Nambu and G.~Jona-Lasinio,
  Phys.\ Rev.\  {\bf 124}, 246 (1961).


\bibitem{DE.rev}
E. J. Copeland, M. Sami,
and S. Tsujikawa, Int. J. Mod. Phys. D 15, 1753 (2006).

\bibitem{SF}
B.~Ratra and P.~J.~E.~Peebles,
  Phys.\ Rev.\  D {\bf 37}, 3406 (1988),C.~Wetterich,
  Astron.\ Astrophys.\  {\bf 301}, 321 (1995)
  [arXiv:hep-th/9408025].

\bibitem{tracker}
Steinhardt,P.J.Wang,L.Zlatev I. Phys.Rev.Lett. 82 (1999) 896, arXiv:astro-ph/9807002;
Phys.Rev.D 59(1999) 123504, arXiv:astro-ph/9812313
9.

\bibitem{quint.ax}
  A.~de la Macorra and G.~Piccinelli,
  Phys.\ Rev.\  D {\bf 61}, 123503 (2000)
  [arXiv:hep-ph/9909459];
  A.~de la Macorra and C.~Stephan-Otto,
  Phys.\ Rev.\  D {\bf 65}, 083520 (2002)
  [arXiv:astro-ph/0110460].

\bibitem{GDE.ax}
  A.~de la Macorra,
  Phys.\ Rev.\  D {\bf 72}, 043508 (2005)
  [arXiv:astro-ph/0409523];
  A.~De la Macorra,
  JHEP {\bf 0301}, 033 (2003)
  [arXiv:hep-ph/0111292];
   A.~de la Macorra and C.~Stephan-Otto,
  Phys.\ Rev.\ Lett.\  {\bf 87}, 271301 (2001)
  [arXiv:astro-ph/0106316];

\bibitem{GDM.ax}
  A.~de la Macorra,
  Phys.\ Lett.\  B {\bf 585}, 17 (2004)
  [arXiv:astro-ph/0212275],A.~de la Macorra,
  Astropart.\ Phys.\  {\bf 33}, 195 (2010)
  [arXiv:0908.0571 [astro-ph.CO]].

\bibitem{IDE}
 S.~Das, P.~S.~Corasaniti and J.~Khoury,Phys.  Rev. D { 73},
083509 (2006), arXiv:astro-ph/0510628;
A.~de la Macorra,  
 Phys.Rev.D76, 027301 (2007), arXiv:astro-ph/0701635

\bibitem{IDE.ax}
  A.~de la Macorra,
  JCAP {\bf 0801}, 030 (2008)
  [arXiv:astro-ph/0703702];
  A.~de la Macorra,
  Astropart.\ Phys.\  {\bf 28}, 196 (2007)
  [arXiv:astro-ph/0702239].

}

\end{document}